# Generation and delivery of free hydroxyl radicals using a remote plasma

H.N. McQuaid[1], D. Rutherford[2], D. Mariotti[1], P.D. Maguire[1]

[1.]*NIBEC, Ulster University, Belfast, BT37 0QB, Northern Ireland,*
[2]*Czech Technical University, Prague, Czech Republic*

**Abstract**

We demonstrate a new gas-based OH$^\bullet$ generation source using a low power RF-driven atmospheric pressure plasma configured to deliver the radical flux into the far effluent region, well away from interference from other plasma factors such as electric fields, currents, and UV radiation. Using He – H$_2$O gas chemistry isolated from the laboratory air, the plasma generated flux consists of H$_2$O$_2$ and OH$^\bullet$, varying with H$_2$O vapour content and absorbed power density. Peak flux values were 2.3 nmol s$^{-1}$ and 0.23 nmol s$^1$ for H$_2$O$_2$ and OH$^\bullet$ respectively at a distance of 50 mm from the plasma, with 790 ppmv H$_2$O and a power density of ~ 10$^8$ W m$^{-3}$. The maximum OH$^\bullet$ flux density was 4.5 x 10$^{19}$ m$^{-2}$ s$^{-1}$ falling to 1.7 x 10$^{19}$ m$^2$ s$^1$ at 110 mm, equivalent to generation rates of 74 µM s$^1$ and 28 µM s$^{-1}$. Despite high OH$^\bullet$ recombination rates at the plasma exit, the escaping flux is still significant, indicating a viable delivery capability to downstream targets. Its performance with regard to OH$^\bullet$ generation rates compares well with traditional OH$^\bullet$ generation techniques such as radiolysis, advanced oxidation processes and enhanced Fenton-chemistry approaches where OH$^\bullet$ production rates are sub-µM s$^{-1}$. Delivering precisely quantifiable OH$^\bullet$ fluxes provides new opportunities for scientific studies and technological opportunities in cell biology, atmospheric chemistry, protein unfolding and systematic dose studies for plasma-based and other OH$^\bullet$ related potential medical treatments.

## 1. Introduction

The hydroxyl radical, OH$^\bullet$, plays an important role in numerous areas including atmospheric chemistry (1), water treatment and pollution remediation (2), antibiotics and disinfection (3), tumour therapy (4,5) and protein studies (6), among others. Fenton chemistry has found wide application as a hydroxyl source, however other areas require alternative approaches to avoid unwanted byproducts, to operate outside low pH environments or achieve much faster reaction times (7). For example, with protein footprinting, direct OH$^\bullet$ generation via synchrotron radiolysis of H$_2$O (8,9) or excimer laser photolysis of H$_2$O$_2$, Chen et al. have demonstrated ultra-fast and irreversible labelling of proteins capable of following their folding/unfolding dynamics (10). Other, more accessible, direct techniques including electrospray and plasma generated OH$^\bullet$ are currently being explored (11). OH$^\bullet$ is considered the most reactive and toxic radical in biology and selective OH$^\bullet$ generation approaches have been proposed for cancer therapy. Gamma radiolysis is known to produce a high concentration of OH$^\bullet$ radicals with strong evidence that these play an important, even dominant role, in DNA destruction and tumour cell death. In enhanced chemodynamic therapy, photodynamic, photothermal, sonodynamic and radiation therapies are used to enhance Fenton-like reactions from nanomedicines in the acidic tumour microenvironment (4,12,13).

Atmospheric pressure plasmas have attracted considerable attention in the past decade as radical sources for application as possible medical treatments for wound healing and cancer (14). These plasma devices are efficient generators of reactive oxygen and nitrogen species (RONS), producing a rich cocktail of species in the gas phase at high concentrations which can be delivered directly to tissue (15,16). Plasmas can also produce high UV photon fluxes, injected currents and electric fields and have been shown to induce significant biological effects (17). Considerable effort has gone into characterising the complex plasma physical and chemical properties on the one hand and attempting to correlate plasma species fluxes with biological outcomes on the other. There are many cold plasma sources designs, and these can generally be classified as either plasma jet (APPJ) or dielectric-barrier discharge (DBD) and can further categorised as either direct-contact or indirect-contact. In the former, the treatment target, including liquid, forms part of the plasma electrical circuit while in the latter, the circuit is isolated, but the plasma is sufficiently close to allow species interaction. Commercially available devices such as the kINPen (14,18) and PlasmaDerm (19) are examples of direct and indirect sources respectively. Irrespective of the electrical circuit, it can be argued that all such plasmas are effectively coupled to the treatment substrate since the plasma-substrate gap is either very small or electric-field propagation via ionisation waves recreates the plasma downstream (20,21). This has the advantage of delivering the high radical fluxes required but brings with it certain technological and scientific disadvantages. The associated electric fields, currents and UV radiation along with potentially high temperature gradients, flow induced turbulence and impurity re-entry into the plasma creates significant challenges for real-time monitoring and responsive control in a clinical environment. From a scientific perspective, the interplay of these multiple synergistic factors represents a major obstacle to systematic plasma – liquid characterisation and simulation, while the inclusion of biological interactions adds an additional layer of complexity that delays our fundamental understanding and predictive capabilities. In this work we aim to investigate a truly non-coupled remote plasma operating at a sufficient stand-off distance from a liquid substrate. By isolating the plasma itself from the ambient environment, and thus excluding air ingress, we have much greater control of the gas chemistry. Using the plasma interaction of helium and water vapour only, we can generate various $HO_x$ (x: 0 - 2) and O species as the primary radicals, along with $H_2O_2$. Other species (e.g. $O_2(a1\ \Delta)$, $O_3$) may be present but at much lower concentrations (27). While $OH^{\cdot}$ flux levels may not match those of direct contact plasmas (22), the relative purity of the radical source and the absence of electric fields and currents may prove to be a major advantage for applications such as tumour therapy or protein footprinting. Also, such as source provides a new tool for investigating individual factors relevant to plasma – biological interactions, allowing thereafter incremental increases in the complexity of plasma chemistry and determination of synergistic or multivariate effects.

Changing the water vapour content of the feed gas has been found to be an effective way to control the $OH^{\bullet}$ radical concentration within the plasma region and therefore in the APPJ effluent itself (23–27). Noble gas plasma interaction with trace water vapour gives rise to many species and reactions due to multiple pathways including electron impact ionisation, excitation, and dissociative attachment as well as ion-molecule reactions. The use of an RF-excited plasma ensures the plasma region is restricted to a short distance (~ mm) from its electrodes (28). In air, the effective rate coefficient for $OH^{\bullet}$ recombination is $1.6 \times 10^8$ $M^{-1} s^{-1}$ with trace (~10 pM) components, e.g., CO, present (29,30). In the plasma effluent, however, the $OH^{\bullet}$ is isolated from ambient air for most of the flight time, and its lifetime is determined primarily by two-body ($OH^{\bullet} + OH^{\bullet} \rightarrow H_2O + O$), ($OH^{\bullet} + HO_2 \rightarrow O_2 + H_2O$), ($OH^{\bullet} + O \rightarrow O_2 + H$) or three-body ($H + OH^{\bullet} + H_2O \rightarrow 2H_2O$) recombination reactions (27). Attri et al. report $OH^{\bullet}$ lifetimes in the gas phase above a liquid surface of ~ 4 ms for high $OH^{\bullet}$ concentration levels around $10^{22}$ $m^{-3}$ and by inference ~ 1 ms for typically reported plasma $OH^{\bullet}$ concentrations (31). This is a directly – coupled plasma source and hence high densities of $H^{\bullet}$ radicals may be expected at the liquid surface which could contribute significantly to $OH^{\bullet}$ recombination at the gas – liquid interface (32). However, with remote plasmas, knowledge of $OH^{\bullet}$ lifetimes and uptake by liquid are not known.

Measurement of neutral and ionic species have been performed in the plasma and afterglow regions via laser-induced fluorescence (LIF) (33,34), UV and VUV absorption (35,36), two-photon LIF (TALIF) (25,37,38), and mass spectrometry (25,39–43) among others. Willems et. al and Benedikt et al. measured the $OH^{\bullet}$ and O species densities in the field-free effluent of a He – $H_2O$ RF (13 MHz) plasma jet, up to 25 mm from the RF plasma source (25,43). Near the plasma, they observed an increase in $OH^{\bullet}$ density ~$10^{20}$ $m^{-3}$ as the $H_2O$ content increased, saturating at ~ 5 000 ppmv. Similar results have been obtained by Bruggeman et al. in the plasma region (44). At 20 mm from the plasma, the $OH^{\bullet}$ density had decreased by over an order of magnitude, for a gas flow of 1.4 SLM and [$H_2O$] > 7 000 ppmv (25) while under similar conditions, $H_2O_2$ concentrations remain relatively constant and $HO_2$ is no longer present. Verreycken et al., with an Ar-$H_2O$ jet into air, observed the reduction in $OH^\bullet$ density by a factor of 10 within 8 mm from the plasma at high $H_2O$ content (33). Reuter et al., using TALIF, determined atomic oxygen density up to 100 mm from the RF He plasma nozzle (45). However, in this case VUV emission from oxygen lines was sufficient to maintain the oxygen density by photo-dissociation, over long distances. Outside the plasma region, dissociative recombination of $H_2O^+$ can be a dominant pathway for $OH^\bullet$ formation, where $H_2O^+$ is often formed by penning ionisation with He metastables (32). In the absence of additional oxygen, loss mechanisms along the jet axis may involve $OH^\bullet$ recombination with $H^\bullet$ radicals to form $H_2O$ and three-body recombination to form $H_2O_2$ (32). The $H_2O + OH^\bullet \rightarrow H_2O + OH^\bullet$ reaction is symmetric and therefore not of significance (46).

For most chemistries, it is difficult to measure gas phase species directly beyond the afterglow / near effluent regions; downstream liquid chemical analysis or simulation therefore have been used to infer upstream plasma chemistry (47–49). With both directly coupled kHz jets (47,49), and remote RF jets (48), there is strong evidence that species ($H_2O_2$, $^\bullet OH$) measured in liquid are created in the plasma gas phase. While liquid chemical probing offers the possibility of determining gas species density profiles upstream, such measurements have not been reported, although Kawasaki et al. obtained relative measurements of ROS variation with distance up to 30 mm, for a directly coupled jet, and Plimpton et al. bubbled plasma ($O_3^-$) effluent through $H_2O_2$ to generate $OH^\bullet$ in liquid far downstream (50,51). Overall, where measurements have tracked changes with distance, such distances are generally < 10 – 20 mm and species densities are observed to drop by a factor of ten. Experimental determination however, of $OH^\bullet$ fluxes and their decay over extended distances from the plasma, has not been carried out.

In this work we present an RF APP operating in helium with a controlled admixture of water vapour for remote radical delivery. We investigate the plasma's ability to deliver reactive molecules to a sample solution up to 110 mm downstream of the plasma by quantifying the flux of $H_2O_2$ and $OH^\bullet$ at the gas/liquid interface. At this distance the sample is far enough removed from the plasma region that any species measured are the result of transport alone and not from plasma interaction with the liquid itself. The effect of feed gas humidity on both the power absorbed by the plasma and $OH^\bullet$ production is reported, and an optimum humidity for maximum $OH^\bullet$ delivery discussed. Specie fluxes were used as inputs to a zero-dimensional kinetics model, modified to simulate the progression of species densities generated in the plasma region into the far effluent. Trends in measured fluxes with changing plasma conditions were compared against the model's output to test the model's validity at predicting long range delivery of $OH^\bullet$ and $H_2O_2$. The performance of the remote plasma source with regard to $OH^\bullet$ production rates is compared with other $OH^\bullet$ generating sources such as radiolysis, advanced oxidation processes and enhanced Fenton chemistry approaches as well as with direct contact and directly coupled plasma devices.

## 2. Methods

The RF APP used in this study (figure 1) consists of a 2 mm inner diameter quartz capillary surrounded coaxially by two 1 mm diameter copper ring electrodes, separated by 2 mm. A 13.56 MHz wave was generated by a Cesar RF power generator and coupled to the electrodes with the aid of an MFJ antennae tuner. A more detailed description and operation of this RF APP device can be found in (52).

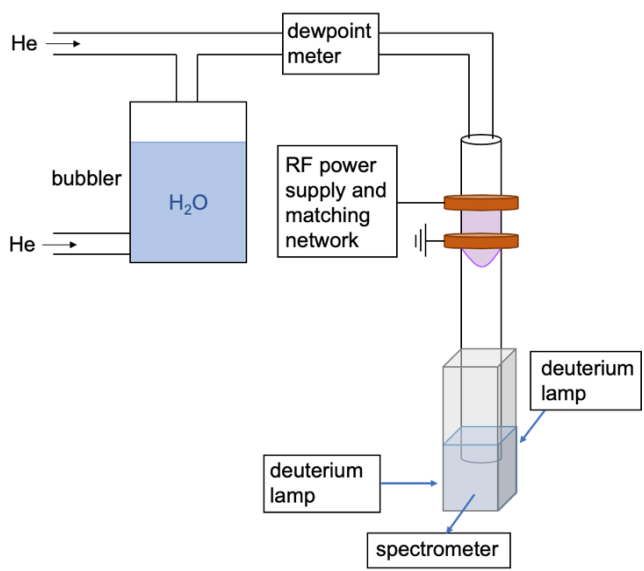

**Figure 1.** Schematic overview of the experimental setup, showing the APPJ and bubbler system for the admixing of water to the helium feed gas.

Helium was fed through the quartz capillary at a mass flow rate of 1.0 SLM and the power absorbed by the plasma device monitored using an inline RF VI probe (Impedans Octiv Suite). The plasma was operated up to a maximum absorbed power of 1.9 W and assuming a cylindrical geometry, the volume was estimated from the length of the luminous region, from which the power density was estimated as ~ 75 MW m$^{-3}$. The visible bulk of the plasma was confined to the interelectrode region with the visible plasma plume extending a maximum of 3 mm beyond the electrodes. To guarantee remote operation, a plasma – liquid distance of > 10 mm provides a sufficient safety margin and at an average gas velocity of > 10 m s$^{-1}$, radical transport times are in the millisecond range. The water content of the plasma feed gas was controlled by mixing two flows; a dry helium flow (CP grade) controlled by one mass flow controller (MFC), and a humidified helium flow controlled by another MFC. The humidified helium was produced by passing dry helium through a water bubbler. Water content could then be regulated by varying the ratio of the dry to humidified helium flows while keeping the total flow at 1.0 SLM. Saturation of the humidified flow was not necessary as the water content was measured using a dewpoint meter (Xentaur LPDT) with an accuracy of ± 3 °C, placed between the mixing point of the two flows and the entrance to the plasma.

$H_2O_2$ concentration was calculated from UV absorbance spectroscopy measurements of titanium (IV) oxysulphate (TiS) solution (Sigma Alrich 495379). A deuterium lamp (Ocean Optics DH-2000-BAL) was used in combination with an Ocean Optics spectrometer (QE65 Pro). TiS was added to the sample cuvette at a concentration of 100 mM, which reacts with $H_2O_2$ to produce pertitanic acid with a characteristic absorbance at 407 nm. Absolute calibration was performed using the absorbance of several samples containing known $H_2O_2$ concentrations. OH• concentration was calculated from fluorescence emission spectroscopy measurements. Sample excitation and emission was measured using the same light source and spectrometer. Terephthalic acid (TA) (Sigma Aldrich 185361), was used at a concentration of 2 mM in 5 mM sodium hydroxide solution. TA reacts with OH• to form 2-hydroxyterephthalic acid (HTA) which emits at 425 nm when excited with UV at 310 nm. While the rate constant for the addition of OH• to TA is around 4.0 x 10$^9$ M$^{-1}$ s$^{-1}$ (53), in $O_2$ containing solutions subsequent reactions produce HTA with a 35 % yield (54,55). OH• flux was therefore calculated by dividing the rate of HTA formation by 35 % and the effective area of OH• delivery to the sample, equal to the cross-sectional area of the capillary. Sample HTA concentration was calculated from calibration data obtained by plotting 2-hydroxyterephthalic (Sigma Aldrich, 752525) concentration against emission intensity. The TA reaction requires $O_2$ which over time will be expelled from the liquid by the He gas flow. We observed a fall in $O_2$ concentration in 5 mL liquid, subjected to 1 slm He flow, from 7 ppmv initially to 0.1 ppmv after two minutes. However, the plasma acts as a source of $O_2$ and from the model we observe at 50mm (figure 6b), our closest measurement distance, that the $O_2$ flux is greater than that of $OH^\bullet$. Since the reaction is catalytic, i.e., $O_2$ is not consumed, sufficient replenishment is obtained from the plasma.

Liquid samples were contained in a quartz micro cuvette, held in place with an Ocean Optics CUV cuvette holder. This setup enabled *in situ* spectroscopy measurements to be taken during plasma treatment. The quartz capillary could be extended distances of 50-110 mm from the lower plasma electrode, and the micro cuvette located at a distance that positioned the exit of the quartz capillary 5 mm beneath the liquid sample surface. Sample volumes were restricted to 0.5 ml; larger volumes resulted in sample ejection from the cuvette during treatment, while smaller volumes failed to provide a bubble free region at the bottom of cuvette for consistent spectroscopy measurements. Because the outlet of the APPJ was positioned beneath the liquid surface with a continuous flow of helium escaping the cuvette, it is reasonable to assume that the plasma system does not contain any significant atmospheric contaminants. In previous surface treatment experiments, a combination of high $OH^\bullet$ concentrations generated near the plasma-liquid interface and limited diffusion, meant absolute calculations of the amount of $OH^\bullet$ transferred from the plasma to the liquid were not possible (56). Bubbling with the plasma effluent produces a homogeneous distribution of $OH^\bullet$ and HTA throughout the sample, enabling more accurate $OH^\bullet$ density calculations. Control measurements consisted of the sample being treated with helium only.

ZDPlasKin (57), a 0D chemical reaction kinetics solver, was used to model the plasma region and downstream species densities. The model used herein accounted for 58 species and 739 reactions that revolved around the He, H and O products produced in a typical $He/H_2O$ plasma. The relevant reactions were taken from a previous study by Aghaei et al. (58), however no N containing reactions were included since high purity He (99.999%) was used throughout, and the capillary outlet is submerged and isolated from air. Rate equations, if not constant, are calculated using electron/gas temperature dependent reaction rates from literature and are integrated in time using the built-in solver. Electron transport coefficients and the rate of electron impact reactions are calculated via Bolsig+, automatically called by ZDPlasKin. See Table 1 in Supplementary Information for the complete list of species and reactions used. Plasma geometry, gas pressure and gas temperature are taken from the experiment and provided as starting conditions. The species concentrations with distance downstream from the plasma i.e., the cathode electrode, is determined from the temporal evolution of the plasma chemistry, once the plasma (ions and electrons) is switched off. Using a constant gas velocity in the capillary, of 5.3 $ms^{-1}$, we convert to a distance response. We define the time period from 0 – 5 x $10^{-4}$ s as the interelectrode plasma region (0 - 3 mm) and 5 x $10^{-4}$ – 2 x $10^{-2}$ s as the downstream region of the ground electrode to the sample interface (3 - 110 mm). Two input parameters were changed in the model at 5 x $10^{-4}$ s. The plasma gas temperature, initially set at 320 K (59), was reduced to the ambient value of 295 K. The OH density at 50 mm is insensitive (< 5% variation) to the downstream temperature between 295 K and 320 K. The plasma voltage was applied up to 5x$10^{-4}$ s and then reduced to zero. The RF voltage magnitude was estimated from in-situ impedance measurements and associated circuit model. All reactions are included for both parts of the model, including electron impact ionisation and excitation reactions. The model outputs of particle density were converted to flux by considering gas velocity and capillary cross-sectional area.

## 3. Results

Plasma treatment was initially performed with a constant water vapour concentration of 600 ppmv with the liquid at a distance 50 mm from the ground electrode (Z = 50 mm). During helium treatment of the controls, HTA concentration increased by 24 %, which coincided with a decrease in volume of ~ 25 %, irrespective of gas composition (helium or plasma effluent). Evaporation of the water solvent alone was therefore identified as the cause, and an evaporation correction factor was therefore applied to all concentration measurements.

See supplementary information S1. The maximum treatment time was 10 minutes, to maintain the liquid surface level above the inserted capillary. At 50 mm from the plasma, $H_2O_2$ density is shown to increase linearly over 5 minutes, reaching a maximum density of $3.7 \times 10^{23}$ m$^{-3}$ (610 µM) for a plasma power of 1.9 W (figure 2a). $H_2O_2$ flux, calculated from the rate of density increase in the cuvette (figure 2a) divided by the capillary cross-section area, also increases with plasma power (figure 2b) from the minimum absorbed power at 0.08 W to a maximum flux of $1.95 \times 10^{20}$ m$^{-2}$ s$^{-1}$ at 1.9 W. At low powers $H_2O_2$ flux increases linearly but starts to saturate at powers higher than 0.54 W. This flux calculation assumes that all incident $H_2O_2$ transfers to the liquid due to the high Henry constant (60).

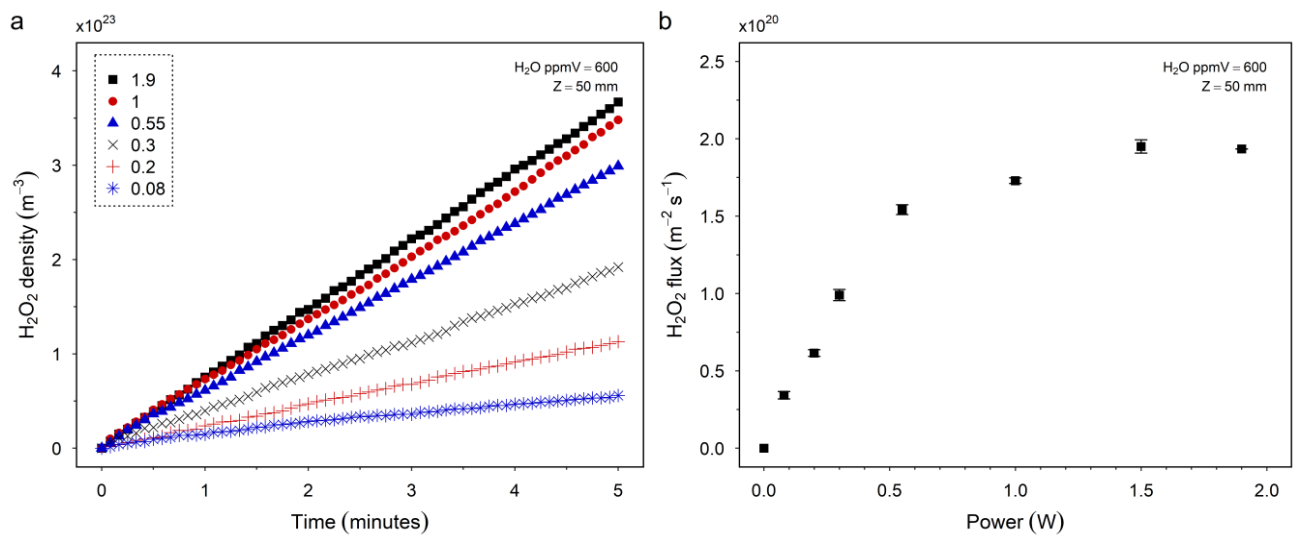

**Figure 2.** Concentration of $H_2O_2$ generated in water containing 100 mM titanium oxysulphate. The sample was located 50 mm downstream and treated using a plasma generated with helium containing $H_2O$ at 600 ppmv. $H_2O_2$ density was calculated in situ from the absorbance of pertitanic acid at 407 nm. (a) Density of $H_2O_2$ produced over time for a range of plasma powers. (b) Average $H_2O_2$ flux density produced over 5 minutes of plasma treatment for increasing plasma power.

Using a similar experimental setup as above, the density of HTA produced in a solution of terephthalic acid was measured over 5 minutes of plasma treatment (figure 3a). HTA density increases almost linearly for the first 3 minutes, after which the rate of HTA increase slows. OH$^\bullet$ flux was calculated using the maximum rate of HTA formation (see methods), in this case over the first 2 minutes. The OH$^\bullet$ flux is shown figure 3b to increase linearly at low absorbed power but tend to saturate at powers > 0.5 W. A maximum flux of $3.4 \times 10^{19}$ m$^{-2}$ s$^{-1}$ is produced at an absorbed power of 1.9 W.

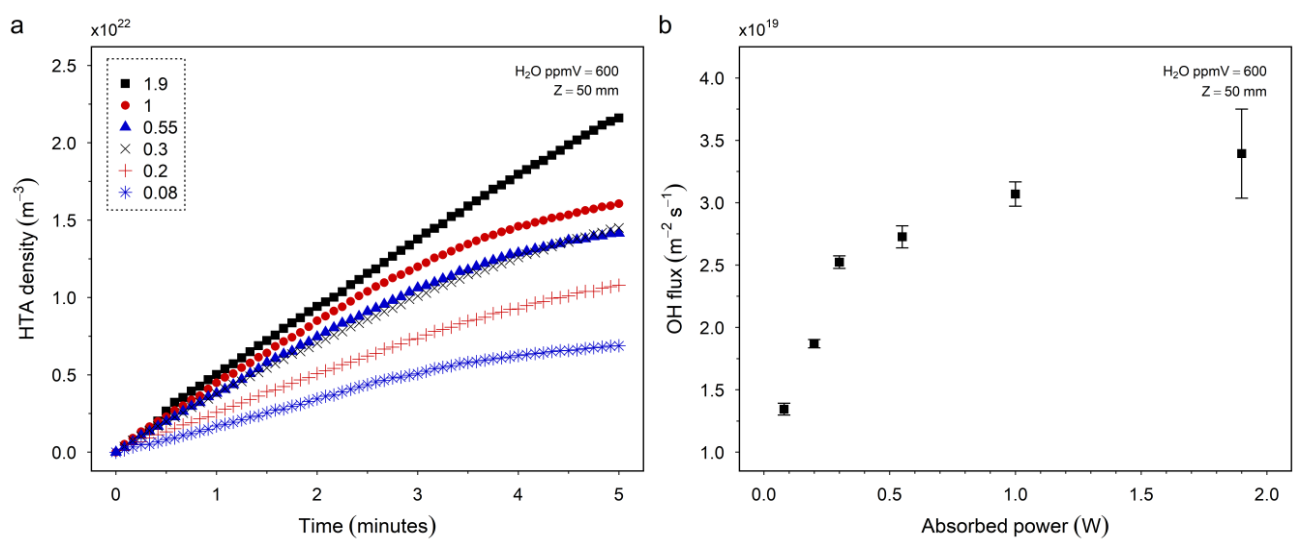

**Figure 3.** OH• density measurements in water containing 2 mM terephthalic acid and 5mM sodium hydroxide. Sample was located 50 mm downstream and treated using a plasma generated with helium containing $H_2O$ at 600 ppmv. OH• density was calculated in situ from the fluorescence emission of 2-hydroxyterephthalic acid at 430 nm. (a) Sample OH• density over time for various plasma powers. Density increased linearly for the first 3 minutes, after which the rate decreased. (b) Maximum OH• flux delivered at the liquid interface over the first two minutes of plasma treatment for increasing plasma absorbed power.

Admixing $H_2O$ to the helium feed gas allowed a maximum plasma operating range up to 4 500 ppmv $H_2O$. The minimum achievable concentration was 7 ppmv in the complete system when operating with pure helium, while above 4 500 ppmv $H_2O$, the plasma became unstable and was extinguished. Keeping the applied power constant, the variation in plasma absorbed power for a given water concentration is shown in figure 4. The maximum absorbed power was 0.77 W at 15 ppmv $H_2O$ and decreases to ~ 0.12 W by 4 400 ppmv, 15 % of the maximum.

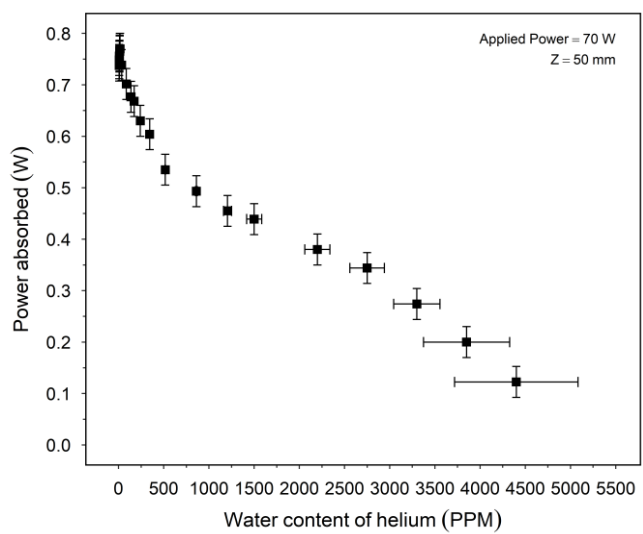

**Figure 4.** Power absorbed by the plasma for varying water content of the helium feed gas. Water content was controlled by admixing a water saturated helium feed to a pure helium feed and measured using a dewpoint meter (Xentaur LPDT). Absorbed power was measured across all $H_2O$ concentrations using an Impedans Octiv Suite 2.0 VI probe, while operating at a constant input power. Horizontal and vertical error bars indicate manufacturers specified accuracy of the dewpoint meter and VI probe respectively.

OH• flux measurements were repeated under the same conditions, at a fixed applied power of 70 W and the variation with $H_2O$ concentration is shown in figure 5a. The initial flux at 7 ppmv of $1.6 \times 10^{19}$ m$^2$ s$^{-1}$ increases steeply, reaching a maximum flux of $2.4 \times 10^{19}$ m$^2$ s$^{-1}$ at 790 ppmv $H_2O$. Above 790 ppmv, the OH• flux decreases for increasing water content, and by 4 400 ppmv $H_2O$, reaches a flux similar to that at < 10 ppmv. The model humidity was set at 600 ppmv and the excitation voltage increased to a value of 395 V (from 0 - $5 \times 10^{-4}$ s) to normalise the density of OH• calculated at 0.001 s (equivalent to 50 mm @ 5.3 m s$^{-1}$) to that measured experimentally at 50 mm. Voltage was then kept constant while water content varied around 600 ppmv. The resulting model flux is plotted alongside the data in figure 5a. OH• flux was predicted to increase slightly as the water content decreased, resulting in a maximum at ~200 ppmv as opposed to the measured maximum at 790 ppmv. OH• flux decreased drastically < 50 ppm, similar to the measured flux, however the model predicted a more abrupt decrease in OH• at higher ppmv. Simulated flux decreased by an order of magnitude by 1700 ppmv, whereas the plasma was still able to deliver a meaningful OH• flux up to at least 3300 ppmv. From the absorbed power characteristic, figure 4, the OH• flux per watt absorbed power is plotted against $H_2O$, figure 5b, and shows a sub-linear relationship at high vapour content. The OH• flux per watt absorbed appears to follow a near √ [H$_2$O] relationship, with the maximum delivered at the maximum H$_2$O concentration sustained by the plasma. A possible physical basis for this relationship is discussed later.

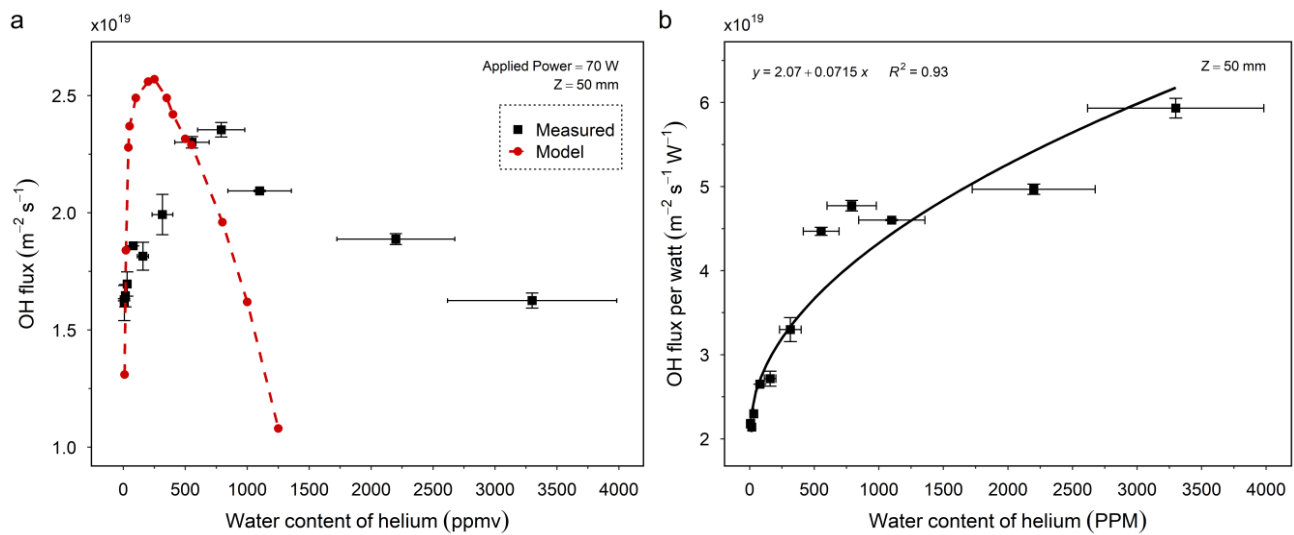

**Figure 5.** (a) Change in the OH$^\bullet$ flux delivery for varying helium feed gas H$_2$O concentration at 70 W applied plasma power. Also shown is the model's predicted variation in OH$^\bullet$ flux with changing H$_2$O content. Both show an ideal water content for maximum OH$^\bullet$ delivery. (b) OH$^\bullet$ flux per watt, calculated by normalising the raw OH$^\bullet$ flux with the absorbed plasma power at each H$_2$O concentration. The trendline indicates a possible OH$^\bullet$ flux per watt $\sqrt{[H_2O]}$ relationship.

Species flux at the liquid interface was measured for increasing distance from the plasma region, while operating at the maximum absorbed power of 1.9 W. Distance of sample from the plasma did not affect the absorbed power. The capillary exit was maintained below the sample surface at all distances to avoid ambient air ingress. Figure 6a shows measurements of OH$^\bullet$ and H$_2$O$_2$ flux values at distances of 50, 70, 90 and 110 mm from the ground electrode. Keeping other parameters constant, the model's plasma region voltage was increased to best fit the fluxes measured at these distances. The model shows a fast decay in OH$^\bullet$ within 2 mm from the plasma edge which is followed by a slower exponential decay over larger distances, similar to that seen experimentally. Modelled H$_2$O$_2$ flux is lower than OH$^\bullet$ inside the plasma region, however, immediately increases outside the plasma and continues to rise slowly with distance.

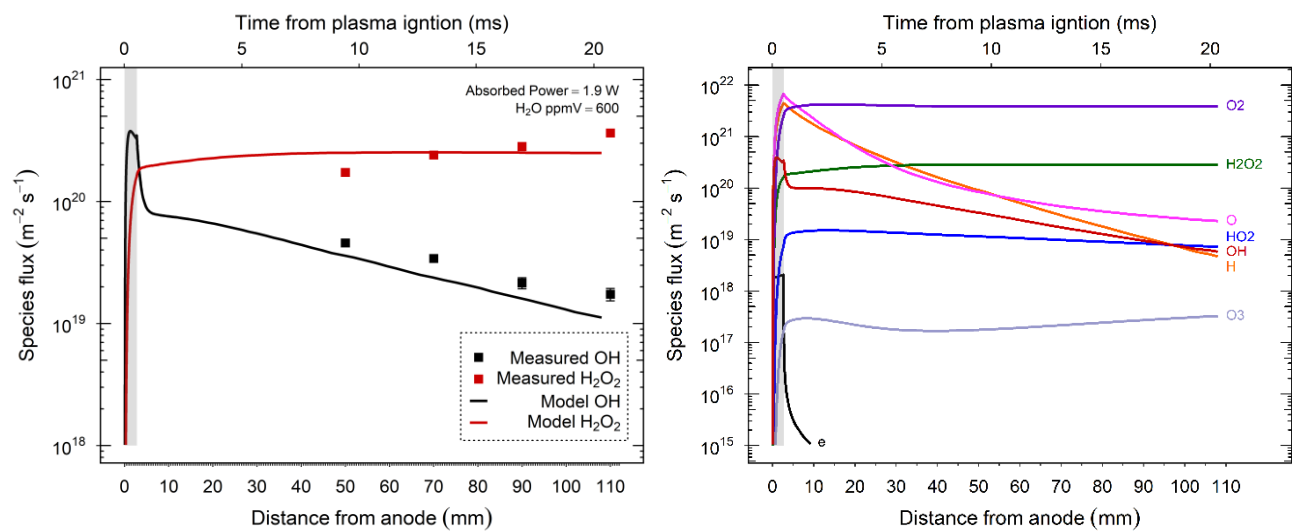

**Figure 6.** (a) Experimental and model results of OH$^\bullet$ and H$_2$O$_2$ densities for increasing distance from the plasma at an initial 600 ppmv H$_2$O content. Model simulates the flux from the start of the plasma region (0 – 3 mm, shaded region) up to a distance of 110 mm. Measurements were taken at 50, 70, 90 and 110 mm from the end of the plasma (ground electrode), with absorbed power kept constant at 1.9 W. Measurements were carried out in triplicate. (b) Flux rates of other reactive species present in the effluent under identical input parameters.

Calibrating the model's flux rates of OH• and $H_2O_2$ against experiment enables us to predict the flux rates of other reactive species present in the effluent. Figure 6b shows that while the flux rates of H and O are much higher than OH• and $HO_2$ at a relatively short distance from the plasma, they also have a faster decay rate over these initial distances.

**Discussion**

Over the distance range 50 mm to 110 mm, set by experimental limitations, we observe a ~ 60 % fall in maximum OH• flux from $4.5 \times 10^{19}$ $m^{-2}$ $s^{-1}$ to $1.7 \times 10^{19}$ $m^{-2}$ $s^{-1}$. Using the model results, a value of $3.2 \times 10^{20}$ $m^{-2}$ $s^{-1}$ is estimated at the plasma edge, equivalent to a density in the plasma of ~ $6.1 \times 10^{19}$ $m^{-3}$. Since the Henry solubility constant for OH• is low, $3.8 \times 10^{-4}$ M $Pa^{-1}$ (60), the liquid surface solubility limit is ~ 20 μM for the equivalent partial pressure. We therefore need to determine if this limit impacts on OH• measurements in the liquid. The time evolution of [OH•] in a shallow surface layer ΔX can be numerically simulated using a simplified flux – reaction model

$$[OH^\bullet](t) = \frac{\Gamma \Delta T}{1000 N_A \Delta X} - \sum_{m=1}^{m=N} k_m [S_m][OH^\bullet] \Delta T \qquad \text{Eq. 1}$$

where Γ is the OH• flux density, $S_m$ the reacting species and $k_m$ the associated rate constants. Considering only the OH• + TA ($k_{OH^\bullet\text{-}TA} = 4 \times 10^9$ $M^{-1}$ $s^{-1}$) and OH• + OH• reactions ($k_{OH^\bullet\text{-}OH^\bullet} = 4 \times 10^9$ $M^{-1}$ $s^{-1}$), the surface OH• concentration reaches a steady-state of ~ 6 μM in less than 1 μs, assuming no surface depletion of the TA probe. The surface layer depth, estimated from the OH• diffusion constant ($\Delta X^2 = D_{OH^\bullet} \Delta T$), is 1.5 nm for ΔT of 1 ns. Under worst-case conditions of static unmixed fluid along with TA depletion, the Henry limit would be reached in ~ 30 ms, with an ultimate steady state OH• concentration of ~ 100 μM. However, since the He gas is bubbled through the fluid, a well-mixed solution is obtained and the Henry limit is unlikely to affect OH• transfer into solution. The Henry limit for $H_2O_2$ is approximately 4 orders of magnitude higher than that for OH•.

Direct comparison with OH• production rates obtained from other OH• generation sources is difficult. Generally, values of OH• concentration in liquid are reported, representing the dynamic balance between generation and loss, whereas flux density values can represent an actual production rate of deliverable species, provided radical losses due to the chemical probe dominate all other reactions, i.e., by using a high reaction rate scavenger at high concentration. Terephthalic acid has a rate constant similar in magnitude to other possible OH• reactions with trace species (OH•, $HO_2$, $O_2^-$) and much higher than OH• - $H_2O_2$ ($3 \times 10^7$ $M^{-1}$ $s^{-1}$). At a concentration of 2 mM, the assumption that the measured TA → HTA conversion represents substantially all OH• reactions in liquid is reasonable. We can therefore convert our observed flux density values to concentration rates, 74 μM $s^{-1}$ (50 mm) and 28 μM $s^{-1}$ (110 mm), for comparison with the non-plasma-based literature. These rates assume flux scales with area, e.g., using extended plasma arrays (the rate would be 0.23 nmol $s^{-1}$ for a plasma jet). In gamma radiolysis, typical dose rates are $10^{-3}$ to $10^{-2}$ Gy $s^{-1}$ (61) and for low LET radiation, the OH• G-value is 2.8 at room temperature, equivalent to a production rate of 0.001 – 0.030 μM $s^{-1}$, while with X-ray radiolysis, production levels of 0.3 μM $s^{-1}$ have been reported (62). In Advanced Oxidation Processes (AOP) with UV excitation, direct measurements of rates (63,64) or of OH• concentrations (65,66) indicate rates up to ~ 0.01 μM $s^{-1}$ whereas with VUV excitation of the gas phase, 70 μM $s^{-1}$ has been demonstrated (67). OH• generation via the traditional Fenton, or Fenton-like reactions, can be enhanced using an external electric field (68) and/or heterogeneous catalysts to deliver rates in the range $0.1 - 1.0$ µM s$^{-1}$ (69,70), while the addition of high concentrations of $H_2O_2$ further increases the rate to $1 - 10$ µM s$^{-1}$(71,72).

For comparison with similar remote RF plasma sources, measured OH$^\bullet$ densities (m$^{-3}$) in the gas phase are often the preferred metric. Gas phase density can be converted to flux density for comparison where gas flow and velocity values are given. Maximum reported flux densities are in the range $2 \times 10^{19} - 2 \times 10^{21}$ m$^{-2}$ s$^{-1}$, with the variability due mainly to the $H_2O$ content, which ranged from 500 ppmv to 11 000 ppmv, and the background gas into which the plasma flows (air or inert) (27,43,44,73–75). The $H_2O$ to OH$^\bullet$ conversion efficiency ($N_{OH}/N_{H2O}$) i.e., the number of OH$^\bullet$ generated per $H_2O$ molecule varies from approximately 1 % to 0.1 %, decreasing with increasing water content. These high flux density values represent conditions measured within the plasma region. However, Schroter et al.(27) observe a 75 % reduction in flux for a gap of 5 mm, with the He plasma effluent isolated from air, while Benedikt et al. also reports a fall of ~ 80 % over 20 mm (25). For argon flowing into an air gap, Li et al. observed ~ 90 % fall over 12 mm (34). The measurements shown in figure 6a, for distances up to 110 mm, represent the first such far effluent measurements of OH$^\bullet$ and show that while a sharp reduction in flux density is likely over the initial 5 mm, thereafter the reduction is much more gradual. Overall, the final flux at 110 mm is ~ 5 % of the mean value at zero gap and by comparison, the remote RF plasma demonstrates a performance capability well above that of other, non-plasma OH$^\bullet$ generation sources, while avoiding the complex additional interactions and species associated with directly coupled plasmas.

We also compared the remote RF OH$^\bullet$ flux densities with those for direct contact or directly coupled plasma sources and configurations. Measurements are either obtained from gas phase, as with the RF plasmas above, or using chemical probes in liquid. In the latter case, for many reported OH$^\bullet$ measurements, only the steady-state OH$^\bullet$ concentration is given, and the information regarding the time to reach steady-state is absent. Low frequency-driven (kHz) plasmas often display a long luminous jet plume due to the propagation of ionisation waves, which maintain the plasma a significant distance beyond the plasma electrodes (76). Gas phase flux densities are reported in the range $5 \times 10^{20} - 10^{22}$ m$^{-2}$ s$^{-1}$, close to the plasma (< 4 mm) and within the plume (33,77,78). Here the plasmas were driven via pin/needle electrodes and electrically coupled to the liquid, while Yonnemori et al, report a lower value of $2 \times 10^{20}$ m$^{-2}$ s$^{-1}$ for insulated DBD electrodes (79). Verreyken et al. report a factor of ten reduction in flux when the gap is increased by 7 mm.

From liquid concentration rates and plasma geometry, we extracted equivalent flux densities. For example, Gorbanev et al. (49) used a parallel field 4 mm diameter plasma jet (18 kV, 25 kHz), with various He – $O_2$ (0 % - 0.5 %) – $H_2O$ (0 % - 4 %) gas mixtures at 4-10 mm from liquid. The liquid remained in contact with the plasma plume and the background environmental gas was He. The maximum OH$^\bullet$ concentration was 24 µM (> 2 000 ppmv $H_2O$, 0 % $O_2$), equivalent to a OH$^\bullet$ delivery rate of 0.4 µM s$^{-1}$. The OH$^\bullet$ concentration was measured by EPR, using DMPO-OH (100 mM) and, with a liquid sample volume of 100 µL the equivalent OH$^\bullet$ flux density is $1.9 \times 10^{18}$ m$^{-2}$ s$^{-1}$. Including 0.2 % $O_2$ with He instead of $H_2O$ reduces the flux by > 40 %, while a further 30 % reduction is observed for the oxygen-water mixture He – $O_2$ (0.2 %) – $H_2O$ (> 2 000 ppmv). Additional OH$^\bullet$ formation due to plasma-induced UV-photolysis directly in liquid was found to be negligible. UV-photolysis rates of ~ 1 pM s$^{-1}$ have also been reported for a similar plasma (1 kV, 35 kHz) with argon (31). Chauvin et al. (10 kV, 10 kHz) report a rate of 0.4 µM s$^{-1}$ and a slightly lower equivalent OH$^\bullet$ flux density, $2 \times 10^{18}$ m$^{-2}$ s$^{-1}$ (80). In this case, the liquid was placed near the limit of the plume and the background gas constituents were He / air (~ 80:20). Uchiyama et al. (18 kV, 20 kHz) obtained $4 \times 10^{19}$ m$^{-2}$ s$^{-1}$ for Ar/air near the edge of the plume but this reduced by an order of magnitude with an additional gap of 10 mm outside the plume (62). These devices utilised isolated DBD electrode configurations while the electrically coupled pin/needle configuration produced a higher flux density, $1 \times 10^{20}$ m$^{-2}$ s$^{-1}$ (81). With larger scale (DBD, streamer or spark) water treatment reactors, typical rates are ≤ 0.05 µM s$^{-1}$, although Kovačević et al. using O$_2$ gas, and a continuous flow water electrode demonstrated maximum OH$^•$ production rates of 12 µM s$^{-1}$ (82–86). The fact that OH$^•$ production rates for the remote non-coupled He – H$_2$O device are similar or higher than directly coupled in-contact plasmas is somewhat surprising since, in the latter, radical generation is expected to be continuous over the whole path from plasma to target. One likely reason is the inevitable presence of air in directly coupled configurations leading to the high-rate production of other radicals (e.g., NO$^•$, NO$_2^•$, NO$_2^-$) which act as OH$^•$ scavengers in the gas and liquid phases (81). Even without air, the presence of high H$^•$ radical concentrations will efficiently recombine with OH$^•$ at the liquid surface (32).

The OH$^•$ radical can be formed by electron impact ionisation, excitation, or attachment of water as well as through metastable – neutral reactions, ion molecule/cluster reactions, ion-ion recombination, and Penning ionisation (24). At lower water concentration, metastable reactions can be important with electron impact dissociation and dissociative electron detachment becoming dominant at ~ 3 000 ppmv (33,34). Assuming a simple loss model with equal OH$^•$ and H$^•$ recombination rates, the OH$^•$ density is expected to follow the density of H$_2$O via a $N_{H_2O}^{0.5}$ relationship, as observed by Bruggeman et al. up to ~1 % H$_2$O (44), although other reports show $N_{OH^•}$ following a linear relationship with $N_{H_2O}$ up to ~ 5000 ppmv before saturating (25,33). Above 6000 ppmv, the OH$^•$ density saturates at 2 - 4 x 10$^{20}$ m$^{-3}$ (25,33,44). These values are similar to the estimated plasma $N_{OH^•}$ in this work of 5 x 10$^{20}$ m$^{-3}$ at 600 ppmv H$_2$O. Regarding peroxide, a linear H$_2$O$_2$ - H$_2$O relationship has been observed (47,48)(43). There are, however, discrepancies in reported concentrations between different measurement techniques (25,43). For example, OH$^•$ density values of 1.5 x 10$^{20}$ m$^{-3}$ have been reported from Cavity Ring Down Spectroscopy (CRDS) and UV absorption, but ~ 6 x 10$^{19}$ m$^{-3}$ from mass spectroscopy at 2000 ppmv. Discrepancies will also reflect the differences in measurement principle that is potentially amplified with instrument location with respect to the plasma source. For example, LIF measurements of OH• at increasing distances from the plasma report increasing discrepancy with those from absorption measurements at the same location, the latter showing almost constant [OH$^•$] with distance up to 9 mm, while LIF indicates an almost one order of magnitude fall (33). However, absorption measurements are line integrated therefore less affected by increasing radial diffusion with distance, hence will more reliably capture the total OH$^•$ flux.

The results show that feed gas humidity plays a fundamental role in the characteristics of the plasma jet used in these experiments. Power absorbed, along with OH$^•$ and H$_2$O$_2$ radical production are all shown to be sensitive to the water precursor content. OH$^•$ production pathways depend on plasma conditions and include via electron dissociation of water or water dissociation products (HO$_2$, H$_2$O$_2$). Other pathways include He metastable interaction with H$_2$O and H$_2$O clusters and excited oxygen or He ion interactions with H$_2$O (27,75,87,88). The model of Schroter et al. (27) of an RF jet indicates electron dissociation of water is dominant at the initial stages of the plasma i.e.

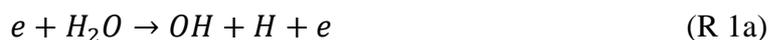
$$e + H_2O \rightarrow OH + H + e \qquad \text{(R 1a)}$$

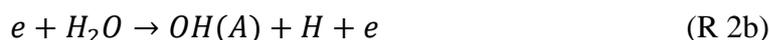
$$e + H_2O \rightarrow OH(A) + H + e \qquad \text{(R 2b)}$$

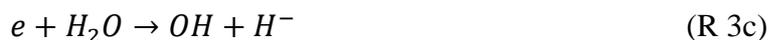
$$e + H_2O \rightarrow OH + H^- \qquad \text{(R 3c)}$$

whereas at later stages, within plasma or afterglow reactions such as H + HO$_2$ → 2OH$^•$ and H + H$_2$O$_2$ → OH$^•$ + H$_2$O can become important. Brisset et al. (75) with a similar jet and simulation model noted the reaction H$_2$O$^+$ + H$_2$O → OH + H$_3$O$^+$ was a significant source of OH$^•$ well within the plasma. By comparison, the plasma device used here is ≤ 3 mm in length, equivalent to the initial stages of (27) where reaction R1 is dominant. It is expected therefore that the OH$^•$ density is largely dependent on a variation in both water content and electron density, which in turn is dependent on power. As previously shown in He – O$_2$ plasmas, an increase in molecular dissociation with absorbed power is a result of a linear increase in electron density with power (38) and a similar trend is therefore also expected for He – $H_2O$. However, above 0.5 W the increase in $OH^\bullet$ is sub-linear (figure 3b), a trend also observed with $H_2O_2$. Since the primary source of $H_2O_2$ is via the recombination of $OH^\bullet$,

$$2OH + He \rightarrow H_2O_2 + He \qquad \text{(R 4)}$$

we expect $H_2O_2$ flux to depend on $[OH^\bullet]^2$ and a log plot of $H_2O_2$ versus $OH^\bullet$ indicates a gradient of 1.95, (figure 7). At powers above 0.5 W, at fixed $H_2O$ concentrations, both $OH^\bullet$ and $H_2O_2$ flux tend to saturate at values of $3.5 \times 10^{19}$ s$^{-1}$ and $1.9 \times 10^{20}$ s$^{-1}$ respectively. Since a change in the $OH^\bullet$ production mechanism is not expected at enhanced electron densities i.e., R1 still holds, the saturation indicates increasing losses, which are consistent with R2.

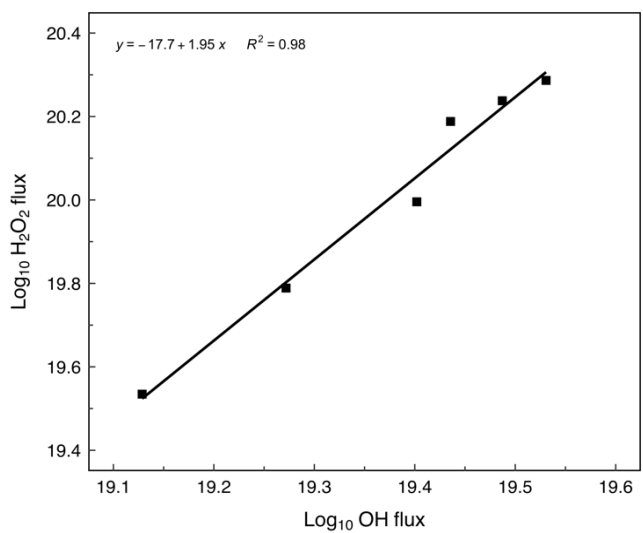

**Figure 7.** Log ($OH^\bullet$) vs log ($H_2O_2$) fluxes under identical operating conditions indicating a gradient of 1.95 close to the expected [$H_2O_2$] dependence on $[OH]^2$.

At fixed applied power with increasing $H_2O$ up to ~ 800 ppm, (figure 5a) the linear increase in $OH^\bullet$ flux is consistent with R1. The significant decrease in absorbed power with $H_2O$ beyond 800 ppmv (figure 4) is likely the main factor leading to the rapid fall in $OH^\bullet$ flux (figure 5a), despite the enhanced precursor concentration at high $H_2O$. For a constant applied power, a reduction in absorbed power with increased $H_2O$ has also been directly observed or inferred by others (25,33). At low $H_2O$ concentrations (< 100 ppmv), the plasma remains electropositive and helium ion species dominate. However as $N_{H2O}$ increases, the plasma becomes more electronegative and the dominant ion species are impurity generated ($OH^+$, $H_2O^+$, $H_3O^+$, $H^+(H_2O)_n$ ion clusters). At > 3000 ppmv, negative ion species, ($OH^-(H_2O)_n$ n = 1-3), become dominant (39). These are formed by dissociative electron attachment to $H_2O$ and subsequent hydration are a significant source of electron loss and a decrease in absorbed power. Also at high water content, an increasing proportion of energy losses into rotational and vibrational excitation and a different electron energy distribution function (EEDF) reduce the rate of dissociation and lead to saturation in $OH^\bullet$ generation. However, from the normalised plot of flux per watt absorbed (figure 5b), the approximate square root relationship indicates that power reduction is not the only factor. This saturation with $H_2O$ was also observed by (25) who noted an increase in the $O/OH^\bullet$ ratio with increasing $H_2O$, the atomic oxygen derived from either $OH^\bullet$ dissociation or recombination. It was suggested therefore that the reaction $OH^\bullet + O \rightarrow O_2 + H$ provides an increasing significant loss channel at higher $H_2O$ concentrations. However, plasma effluent chemistry may be significantly different from that in the near plasma region as short-lived and charged species recombine rapidly at the plasma edge and long-range transport is dominated by more stable neutral species such as $HO_2$ and $H_2O_2$. Simulations (27) in the near-plasma afterglow region indicate the main consumption pathway of $OH^\bullet$ is via its reaction with $H_2O_2$,

$$OH + H_2O_2 \rightarrow HO_2 + H_2O \tag{R 5}$$

It is therefore more likely that the higher densities of $H_2O_2$ produced at higher powers are the primary loss channel for $OH^\bullet$ radicals over the 50 mm path length from plasma to collection liquid. The main $H_2O_2$ generation mechanism in the gas phase is via $OH^\bullet$ three-body recombination, R2 ($OH^\bullet + OH^\bullet + M \rightarrow H_2O_2 + M$) (47).

Combining the generation and loss reactions from R1 and R2 we can estimate the steady state $OH^\bullet$ concentration as

$$[OH^\bullet] = \sqrt{\frac{k_1}{k_2} \frac{[n_e]}{[He]}} \sqrt{[H_2O]} = k^* \sqrt{[H_2O]} \qquad \text{Eq. 2}$$

where $k_1$ and $k_2$ are the rate constants for R1 and R2 respectively, and the slope $k^*$, can be obtained from figure 5b, for 50 mm distance. To obtain the equivalent slope in the plasma region, the flux density at the plasma edge is taken from figure 6a, and the ratio of fluxes at 0 mm and 50 mm is used as a scaling factor, leading to a value of $k^*$ (0 mm) of $2.7 \times 10^{18}$ m$^{-3/2}$. See Supplementary Information S2. After compensating for a difference in power densities, comparison with a similar characteristic by Bruggeman et al., gives a $k^*$ value ratio equal to ~ 0.5 (44), which is commensurate with the difference in electron density between the two systems (44)

The model simulation proved reliable in predicting the $OH^\bullet$ flux characteristics of the plasma effluent. A maximum flux delivery was predicted at ~ 200 ppmv, similar to that seen experimentally at 790 ppmv. While previous studies have shown that $OH^\bullet$ density in the plasma region continues to increase with humidity (25,33,44), the data here shows that trends in plasma density are not necessarily an indication of trends in the effluent. The model proved accurate at predicting both the $OH^-$ and $H_2O_2$ fluxes over extended distances. This enabled the flux of other reactive species in the effluent to be estimated with reasonable reliability. As evidenced by the model, while high flux delivery of $H_2O_2$ up to 110 mm was expected due to its long lifetime, the effluent also delivers multiple highly reactive species, such as $OH^\bullet$, H and O over long distances.

**Conclusions**

We have demonstrated a new gas-based $OH^\bullet$ generation source using a low power RF-driven atmospheric pressure plasma configured to deliver the radical flux into the far effluent region, well away from interference from other plasma factors such as electric fields, currents and UV radiation. Using He – $H_2O$ gas chemistry isolated from the laboratory air, the effluent flux consists of $H_2O_2$ and $OH^\bullet$ with measured flux values of 2.3 nmol s$^{-1}$ and 0.23 nmol s$^{-1}$ respectively at a distance of 50 mm from the plasma. The $OH^\bullet$ flux density was $4.5 \times 10^{19}$ m$^{-2}$ s$^{-1}$ falling to $1.7 \times 10^{19}$ m$^{-2}$ s$^{-1}$ at 110 mm, equivalent to generation rates of 74 µM s$^{-1}$ and 28 µM s$^{-1}$. Millimeter scale RF-driven plasmas are an ideal configuration for remote delivery of selected species as the plasma itself is restricted to mainly within the electrode region, rather than extending well beyond the electrodes in a long jet, as found with lower frequency plasma configurations. It is well known that short-lived radicals suffer very high recombination rates within a few millimeters of the RF plasma exit. Despite this, the escaping flux is still significant, indicating a viable delivery capability to downstream targets. Its performance with regard to $OH^\bullet$ generation rates compares well with traditional $OH^\bullet$ generation techniques such as radiolysis, advanced oxidation processes and enhanced Fenton-chemistry approaches where rates are sub µM s$^{-1}$ and the potential therefore exists for a large enhancement of $OH^\bullet$ supply using RF plasma arrays. Alternatively, the use of single plasma devices delivering precisely quantifiable $OH^\bullet$ fluxes provides new opportunities for scientific studies in cell biology, atmospheric chemistry, protein unfolding and systematic dose studies for plasma-based and other OH• related medical treatments. The potential also exists for OH• and other radical transport over longer distances in fixed or flexible tubing, opening the possibility for direct in-vivo studies leading to new radical-based treatments and tumour therapies.

## Acknowledgements

This work was supported by Engineering and Physical Sciences Research Council (Project Nos. EP/K006088/1, EP/K006142/1, EP/K022237/1, EP/R008841/1, EP/T016000/1) and EU COST Actions PlAgri (CA19110) and PlasTHER (CA20114).